\documentclass[journal]{IEEEtran}

\usepackage{ifpdf}
\usepackage{cite}

\ifCLASSINFOpdf
    \usepackage[pdftex]{graphicx}
    \DeclareGraphicsExtensions{.pdf,.jpeg,.png}
\else
    \usepackage[dvips]{graphicx}
    \DeclareGraphicsExtensions{.eps}
\fi

\usepackage{amsmath}
\usepackage{amsfonts}
\ifCLASSOPTIONcompsoc
  \usepackage[caption=false,font=normalsize,labelfont=sf,textfont=sf]{subfig}
\else
  \usepackage[caption=false,font=footnotesize]{subfig}
\fi

\usepackage{algorithmic}
\usepackage{array}

\usepackage{booktabs}
\usepackage{multirow}

\usepackage{url}
\usepackage[hidelinks]{hyperref}

\usepackage[finalnew]{trackchanges}
\addeditor{IE}
\addeditor{WJ}
\addeditor{DW}

\begin{document}

\title{Survey of Privacy-Preserving Collaborative Filtering}

\author{Islam~Elnabarawy,~\IEEEmembership{Student Member,~IEEE,}
        Wei~Jiang,~\IEEEmembership{Member,~IEEE,}
        and~Donald~C.~Wunsch~II,~\IEEEmembership{Fellow,~IEEE}%
    \thanks{
		I. Elnabarawy is with the Department of Computer Science, Missouri University of Science and Technology, Rolla, Missouri 65409. E-mail: elnabarawy@ieee.org}%
    \thanks{
        W. Jiang is with the Department of Electrical Engineering and Computer Science in the College of Engineering and the Management Department in the College of Business at the University of Missouri-Columbia, Columbia, Missouri 65211. E-mail: wjiang@missouri.edu}%
    \thanks{
        D. Wunsch is with the Department of Electrical and Computer Engineering, Missouri University of Science and Technology, Rolla, Missouri 65409. E-mail: dwunsch@ieee.org.}%
    \thanks{
        I. Elnabarawy and D. Wunsch are affiliated with the Applied Computational Intelligence Laboratory, Missouri University of Science and Technology, Rolla, Missouri 65409. Website: acil.mst.edu}%
}


\maketitle

\begin{abstract}
    Collaborative filtering recommendation systems provide recommendations to users based on their own past preferences, as well as those of other users who share similar interests. The use of recommendation systems has grown widely in recent years, helping people choose which movies to watch, books to read, and items to buy. However, users are often concerned about their privacy when using such systems, and many users are reluctant to provide accurate information to most online services. Privacy-preserving collaborative filtering recommendation systems aim to provide users with accurate recommendations while maintaining certain guarantees about the privacy of their data. This survey examines the recent literature in privacy-preserving collaborative filtering, providing a broad perspective of the field and classifying the key contributions in the literature using two different criteria: the type of vulnerability they address and the type of approach they use to solve it.
\end{abstract}



\section{Introduction}\label{sec:introduction}

The use of recommendation systems has grown significantly in recent years. Shopping websites present users with item recommendations based on their history and demographics; movie and book recommendation websites are being used every day to pick new favorites; and online music streaming services generate dynamic playlists to suit each user's preferences.

Collaborative filtering (CF) is a popular and successful approach to providing user recommendations using knowledge about the user's preferences and the preferences of other users with similar interests to predict which items the user is most likely to be interested in \cite{Adomavicius2005,Su2009}. There is a large body of literature on CF algorithms \cite{Adomavicius2005,Breese1998,Das2007,Castro2007,Castro2007a,George2005,Herlocker2004,1167344,4633869,Resnick1994,Su2009,Zhu2009}, which can be categorized into memory-based and model-based techniques \cite{Adomavicius2005,Su2009}.

Since the key point in CF systems relies on users' preferences and past actions to make predictions, many users may feel uneasy because of privacy concerns \cite{6686270,Calandrino2011231}. Additionally, the need may arise for two or more CF systems to leverage their combined data to provide their users with more accurate recommendations. This type of computation that relies on data from more than one party is referred to as multi-party computation \cite{goldreich1998secure}. To leverage this type of computation, the parties need to be able to do it securely, without allowing the other parties to read and store the user data. This is when privacy-preserving data mining techniques \cite{Agrawal2000} become necessary, leading to many different algorithms for privacy-preserving CF (PPCF); e.g: \cite{Ahmad2007,1004361,5066586,4403128,6570605,6133147,7033684,6785787}.

This survey examines a large number of contributions to PPCF recommendation systems in published literature. It divides the papers into broad categories multiple times based on different factors, with the goal of providing a comprehensive overview of the recent literature in the field. This is a broad perspective that the authors believe to be missing from the recent PPCF literature. The remainder of this survey is organized as follows: Section~\ref{sec:background} provides some background information on privacy-preserving data mining (Section~\ref{ssec:privacy-preserving}) and CF recommendation systems (Section~\ref{ssec:collaborative-filtering}). Section~\ref{sec:survey} contains a survey of the recent PPCF literature, organized into subsections corresponding to the different classifications. Finally, Section~\ref{sec:conclusion} concludes the survey with a summary of the researchers' observations after examining the recent literature.

\section{Background}\label{sec:background}

\subsection{Common Privacy-Preserving Techniques}
\label{ssec:privacy-preserving}

\begin{itemize}
	\item \emph{Secure Multiparty Computation}\\
	To maximize
	privacy or minimize information disclosure, Secure Multiparty Computation (SMC) is the goto technique which
	was first introduced by Yao's Millionaire problem \cite{Yao86}.
	This was extended to multiparty computations by
	Goldreich et al. \cite{Goldreich87}. SMC can be categorized as either information theoretic 
	or computational \cite{BGW88}. In the computational model,
	the adversary is assumed to be bounded by polynomial-time. In the information theoretic model,
	the adversary is assumed to be unbounded. Much work exists to address various aspects (e.g.,
	complexity, adversarial behaviors, the number of corrupted parties) of SMC. 
	There are generally two types of adversaries related to the SMC definitions: semi-honest and malicious \cite{Goldreich}. 
	The semi-honest adversarial model often leads to more efficient privacy-preserving protocols, but the malicious model is less restrictive and thus more realistic. 
	\item \emph{Randomization and Perturbation} \\
	The main idea in the perturbation approach is that data are modified (e.g., adding noise but preserving
	the underlying statistics) before being disclosed and analyzed. 
	The key is that the original distribution can be 
	roughly reconstructed from the perturbed data.
	The paper by Agrawal and Srikant \cite{Agrawal00} introduced this notion.
	In addition to additive noise, multiplicative noise can also be used \cite{LiuTKDE06}.
	
	\item \emph{$k$-Anonymity} \\
	The technique was developed to prevent external linking attacks \cite{kanonymity,kanon_defn}. The basic idea is that 
	a dataset is $k$-anonymous if each record appears at least $k$ times according to 
	a pre-defined set of quasi-identifier attributes. 
	The main approach to achieve $k$-anonymity is information generalization and suppression. 
	
\end{itemize}

\subsection{Collaborative Filtering}\label{ssec:collaborative-filtering}

CF addresses the problem of finding the degree to which a given user would like an item based on the knowledge of what similar users thought of that item and other similar items \cite{Su2009}. It is primarily based on the assumption that users who share similar interests for a subset of items will share the same opinion for other items \cite{Ahmad2007,Adomavicius2005}.

In formal terms, the system is given a set of $N$ items $$\mathbb{X} = \{x_i: i \in 1,2,...,N\},$$ a set of $M$ users $$\mathbb{U} = \{u_j: j \in 1,2,...,M\},$$ and a set of ratings $r_{i,j}$, where $r_{i,j}$ corresponds with the rating that item $x_i$ received from user $u_j$. Given a set of known ratings, CF attempts to find the predicted rating $\tilde{r}_{t,k}$ that user $u_k$ is most likely to assign to item $x_t$. 


The range of values for the ratings varies by domain. Some domains only consider the explicit ratings that the users make, either by assigning a rating to an item or by making a purchase through the system \cite{1167344}. Other domains that lack access to this type of information often rely on other metrics, such as the number of user clicks on web pages and the relative time the user spends looking at a certain web page \cite{Das2007}.

CF techniques face many challenges, such as dealing with sparse rating datasets and large volumes of user and item data. They are often required to generate recommendations in real-time or with near real-time requirements, and incorporate new ratings from users while the system is in operation. Additionally, rating data is often noisy and subjective, which may lead to inaccurate recommendations \cite{Su2009}.

One of the main challenges for CF systems is maintaining the privacy of user information. Various attacks can be used on recommendation systems to expose the preferences and behavioral patterns of a specific user or identify a user within a particular dataset \cite{MacAonghusa2015}\note[IE]{This paragraph is too short and may be out of place.}.

CF approaches can be classified into two categories; memory-based and model-based. There are also hybrid techniques that incorporate both the memory-based and model-based approaches \cite{Adomavicius2005,Su2009}\note[IE]{This paragraph is too short.}.

In memory-based CF techniques, a heuristic is used to make rating predictions based on some or all of the ratings that are already known \cite{Adomavicius2005,Su2009}. They often rely on identifying users that share similar interests with a given user, and use the known ratings of those users to make predictions about the preferences of the user.

Different memory-based CF approaches employ different measures to compute the similarity between users \annote[IE]{and how they make their predictions}{awkward and unclear...}. For example, GroupLens, one of the early generation CF systems \cite{Resnick1994} used the Pearson correlation measure to find similar users, then made its prediction using an average of the ratings of similar users for the same item, weighted by the absolute value of the correlation between each of those users and the user in question.

Model-based CF techniques learn a model based on the known ratings, and use that model to predict how a user would rate a previously-unrated item \cite{Adomavicius2005,Su2009}. The learning system uses the training data to build a  model that represents the patterns and relationships in the training data, and then it uses that model to make predictions about the preferences of the users in the system.

Model-based CF systems vary based on the type of data they are analyzing. For categorical ratings, classification-based algorithms are often used to build the CF model. Conversely, regression models are often used when the ratings are numerical or continuous. An example of some early work on model-based CF \cite{Breese1998} used Bayesian models, where a probabilistic Bayesian network is built using the known ratings, and  used later to predict the value for an unknown rating.
\note[IE]{Add a sentence or two about Matrix Factorization, and cite our paper.}

Hybrid CF techniques join multiple approaches to build a more flexible and robust recommendation system. They often rely on domain knowledge and heuristics to take advantage of multiple aspects of the problem and provide more accurate and personalized recommendations. They may rely on item content, user demographic information, or combine both memory-based and model-based approaches to build the CF system \cite{Su2009}. Some work in the literature, such as \cite{Das2007}, joins both memory-based and model-based techniques into hybrid techniques. These aim at leveraging some details about the problem domain to build systems that bridge the gap between memory-based and model-based techniques, thus providing an added advantage in cases where a new user or item is being introduced to the system.

The interested reader is referred to \cite{Su2009} and \cite{Karydi2016}, which provide surveys of the key developments to the different CF approaches as well as an evaluation of the advantages and disadvantages of each approach.

\subsection{Related Studies}\label{ssec:related-work}

There have been some recent studies with similar goals that examined recent PPCF literature. In one of the more recent studies \cite{Ozturk2015276}, the authors looked at the current trends in PPCF and considered possible future trends for the field. The study looks at other recent survey papers, examines their strengths and shortcomings and then classifies the recent papers in the field with respect to the trends common between them and the different aspects of the CF algorithms used. The authors provided an analysis of the number of publications in the field over the past two decades, and identified three main goals that they think future research should address simultaneously: privacy, accuracy and online performance.

The authors of \cite{BILGE2013Survey} provided a comprehensive survey of the PPCF literature until 2013. They used several different attributes to classify the PPCF literature, including data partitioning cases and the techniques they used to preserve privacy, and presented guidelines and potential future directions for research in the PPCF field. It is one of the most comprehensive survey papers in recent years.

The study by \cite{Bilge2013a} provides a thorough comparison between different clustering-based PPCF approaches. It evaluates the capabilities of recent literature in clustering-based CF in terms of privacy, and provides ways to apply different families of clustering algorithms to the CF problem while keeping the user information private.

In \cite{6686270}, the authors classify the recent PPCF literature into two categories based on the centrality of data. They consider centralized methods to be ones in which the data is stored in one central location and decentralized methods to be ones where the data is distributed between multiple parties, or where users maintain control of their own rating profile. They discuss the current trends based on their survey of the literature and list a number of current PPCF challenges as well as new ones that they anticipate.

Other recent studies include \cite{Jeckmans2013263} and \cite{Batmaz2017}, which provide a more general examination of privacy in recommendation systems, as well as \cite{Gunes2012} and \cite{Lam2004}, which focus on one specific vulnerability known as shilling attacks.

\section{Survey of Recent Work}\label{sec:survey}

Due to the popularity of CF systems, and the prevalence of privacy concerns among users and developers of CF recommendation systems, there have been many contributions to this area of the literature in recent years. The study presented here takes a different approach that was not found in recent surveys on PPCF. It examines and classifies the papers in the field multiple times into broad categories, based on different factors. First, it highlights the way authors of different papers defined privacy and identifies the problem each paper was trying to solve. Then, the papers were classified with respect to how they solved the general PPCF problem, breaking them up into multiple categories corresponding to the most commonly used approaches in the literature.  This classification is meant to provide a broad perspective of the field and help researchers and practitioners interested in PPCF identify the approach that is best suited to their needs.

\subsection{Vulnerability}

Online CF recommendation systems can be subject to different vulnerabilities that compromise the privacy of the users' data and the overall integrity of the system. In this section, the recent PPCF literature is examined and the type of vulnerability that each study attempts to address is identified.

\subsubsection{User Profile Exposure}

Most of the recent PPCF literature attempts to address the general problem of user profile exposure by providing guarantees that the users' preferences remain private and are not exposed to other system users or system admins or being sold to marketing agencies. Table~\ref{tab:vulnerability} cites examples from recent PPCF literature where the studies focused on solving this problem.

The definition of the profile exposure vulnerability often varies, but the common factor used to identify papers addressing this vulnerability is that they focus on protecting the user profile data before it is stored or transmitted, using obfuscation, microaggregation or cryptographic approaches. Sometimes the objective is to protect the data from semi-honest third parties, while other times it is providing the user with guarantees that their profile data is safe from prying eyes, even within the domain of the service providing the CF recommendations.

\subsubsection{Inference Attacks}

The second most popular problem addressed by PPCF literature is that of inference attacks. In this category, the main challenge is protecting against malicious system users who try to inject fake profiles into the system or collect a lot of recommendation data, with the goal of inferring or identifying profiles of other users in the system. This can apply to either a semi-honest or a malicious attack model, depending on the level of security guarantee that the system provides. In contrast with user profile exposure, the system can have access to the complete user profile data, and the objective is to prevent other parties or system users from inferring information about that data. Examples of this category are cited in Table~\ref{tab:vulnerability}.

Studies that address vulnerability to inference attacks often rely on providing $k$-anonymity guarantees to user profiles such that a user profile cannot be identified within a group of $k$ profiles. They may rely on data obfuscation or clustering approaches, as well as microaggregation, to provide these guarantees. This is a more challenging problem in the area of PPCF, where a trade-off may be necessary between the accuracy of the provided recommendations and the degree of anonymity that the system provides.

\subsubsection{Shilling Attacks}

Another popular vulnerability that PPCF studies attempt to address is shilling attacks. This is an attack in which a malicious entity can create a large number of fake profiles that have false ratings with the goal of influencing the system towards or away from recommending certain items. The last row of Table~\ref{tab:vulnerability} cites examples of this type of study.

The study by \cite{Bilge20143671} discusses the sensitivity of different CF schemes to shilling attacks, and suggests that clustering-based CF methods are more robust and can detect and exclude fake profiles. Although this vulnerability may not influence the privacy of user data directly, it can greatly bias a CF recommendation system and decrease its accuracy and usability significantly.

\begin{table}[h]
  \centering
  \caption{Papers divided by vulnerability}
  \label{tab:vulnerability}
  \begin{tabular}{ll}
    \toprule
    \textbf{Type of Vulnerability} & \textbf{Papers} \\
    \midrule
    User Profile Exposure & \cite{Ahmad2007, 5616200, Aimeur2008, Amatriain2009, Armknecht201165, 6483317, baraglia2010preserving, Baraglia2006559, Basu20141706, Basu2012447, 6133147, 6676721, Basu2012, BILGE2013Survey, Bilge2010, Bilge2013a, Bilge2012, Bilge2013, Pathak2013, Borole20141364, Boutet2014169, 1004361, Castagnos2006, Castagnos2008247, 6406488, Erkin201361, Erkin201277, Jeckmans2014, Jeckmans2013400, 6261088, Kaleli2012, Kaleli2010, Kaleli201547, 6332047, 5066586, Kikuchi2013617, Li2016311, Li2011, Liu2015781, Luo2009, 6615356, Memis2013166, 6570605, 7217100, Montaner2002, 6680862, Okkalioglu2016199, 4403128, Polat2005a, 1250993, Polat2005b, Polatidis201562, Scipioni20112011, Shang2013, Smirnov2015455, 5474755, 7044660, Boutet2016827, Troiano201456, 6890141, Wang2013231, Wu2016144, Yakut2011, Zhang2014, 6682707, 7095527}       \\
    \midrule
    Inference Attacks     & \cite{MacAonghusa2015, Calandrino2011231, Canny2002, Casino2013490, Casino20151000, Chen2014218, Clemente2015725, Frey20155, Gao2014717, Honda201539, Honda201243, 7033684, Ramakrishnan2013, Shao2014244, Shokri2009, Wang2014224, Weinsberg:2012:BIO:2365952.2365989, Zhang2014320, Zhang2006, Zhao2015, 7195575, 6785787, Zhu2014}       \\
    \midrule
    Shilling Attacks      & \cite{Bilge20143671, bilge2013robust, Chirita2005, Gunes201513, Gunes201354, Gunes2012, Gunes20161, Lam2004, Okkalioglu2015901}      \\
    \bottomrule
  \end{tabular}
\end{table}

\subsection{Approaches to Privacy}

This survey's authors used their observations of the recent literature to divide the contributions into several categories below based on their approach to preserving privacy: cryptography-based, obfuscation-based, clustering-based, and heuristic-based techniques. 

Cryptography-based techniques (\ref{ssec:cryptography}) rely on using cryptography to secure the user data and often use homomorphic encryption to allow computations to be applied to the data securely. Obfuscation-based techniques (\ref{ssec:obfuscation}), on the other hand, rely on applying transformations to the data to obfuscate or anonymize it, such that the individual users cannot be identified within the data or results. Clustering-based techniques (\ref{ssec:clustering}) use clustering to group users together into small communities, then extract features that are representative of this community as a whole, and use those to generate the recommendations. Therefore, this guarantees that no individual user's data will be identified. Finally, other techniques in the literature (\ref{ssec:heuristic}) rely on alternate methods, such as generating recommendations based on items instead of users, to generate the recommendations without exposing the user data.

In the following subsections, some of the recent work is categorized in terms of the approach to preserving privacy, and some potential advantages and disadvantages to each type of approach are examined. 



\subsubsection{Cryptography-based techniques}\label{ssec:cryptography}

Cryptography-based PPCF appears to be the most popular category in recent literature. Techniques in this group rely on cryptographic measures to carry out the calculations needed for providing recommendations securely, without compromising the privacy of the users' data. They are often used to prevent the user profile exposure vulnerability discussed earlier, which most commonly occurs under a semi-honest attack model.

One of the earlier contributions to this area was by Canny \cite{1004361,Canny2002}. In \cite{1004361}, a cryptographic algorithm for computing a public aggregate of the data was presented that can then be used to securely generate personalized recommendations for individuals. It uses homomorphic encryption to apply the CF calculations on the data and decrypt the results without exposing the individual users' data. The work in \cite{Canny2002} extends the idea further by creating a method based on an Expectation Maximization probabilistic factor analysis model and using a privacy-preserving peer-to-peer homomorphic encryption protocol to apply the CF calculations. The interested reader is referred to \cite{1004361} and \cite{Canny2002} for a complete exposition of the approach summarized here.

In \cite{Ahmad2007}, the authors present an architecture for PPCF using the notion of distributed trust, which relies on a coalition of trusted servers instead of a single server. This distribution of trust among multiple servers makes the system more resilient against faults and attacks while providing an element of privacy for the users' data. The implementation of this architecture relies on a threshold homomorphic encryption protocol and was implemented and evaluated in an experimental setting.

The work presented in \cite{5066586} attempts to address one of the drawbacks of cryptography-based techniques, which is the high computational overhead of encrypting and decrypting large amounts of data. In this contribution, the authors rely on clustering the items and sampling the user data to reduce the computational burden of the cryptographic protocol. The reduced data is then used in a homomorphic cryptography scheme to securely generate CF recommendations with a significant reduction in computational time. In \cite{6332047}, the authors continue to address the goal of improving the time performance of cryptographic PPCF by introducing a quasi-homomorphic similarity measure that allows the use of local similarities to approximate the global similarity and analyze the accuracy of this approximation approach in addition to its running time improvement.

The contributions in \cite{6133147} and \cite{6676721} discuss some important practical considerations for implementing CF on cloud platforms, privacy and security being among the chief concerns of implementing such systems. In \cite{6133147}, they present a practical implementation of a PPCF system, based on the Google App Engine for Java (GAE/J) cloud platform. They designed algorithms that rely on a homomorphic encryption scheme to preserve the privacy of user data in the cloud. This work is further analyzed and extended in \cite{6676721} to address real world Software-as-a-service and Platform-as-a-service cloud settings.

Some of the more recent contributions addressed more specific concerns, such as horizontally-partitioned datasets \cite{6261088}, overlapped ratings \cite{6570605}, and updating the user preferences in real time \cite{7217100}. Others, such as \cite{6890141} and \cite{7044660}, designed cryptography-based PPCF algorithms for other tasks \cite{6890141} or incorporated cryptographic methods with other approaches to create efficient PPCF systems \cite{7044660}. 

While cryptography-based techniques have the advantage of providing reliable security without sacrificing the accuracy of their results, one of the main concerns for this family of techniques is the scalability to systems that require the processing of millions of items and users, especially in online settings where the response time needs to be minimal while providing accurate recommendations.

\subsubsection{Obfuscation-based techniques}\label{ssec:obfuscation}

In obfuscation-based techniques, user profile data is transformed in some way that prevents individual users from being identified by using the data or the system's output, while maintaining the same or a close level of accuracy in the generated recommendations. This is often used to guard against inference attacks by other system users or external entities, following either a semi-honest or a malicious attack model.

The work introduced in \cite{1250993} relies on randomized perturbation techniques to introduce randomness in the data such that a user could not be individually identified with any certainty. Similarly, \cite{Kikuchi2013617} applies some randomness to the response provided by the system to prevent the preferences of individual users from being inferred from the system's responses while allowing the users to calculate the exact response using the randomized one.

Another popular approach in this category is based on the obfuscation of user data, and was examined by \cite{4403128, 7033684, 7195575}. Similar to perturbation, the relevant fields in user data are obfuscated to mask any identifying information in the data, preventing anyone from inferring the original users' preferences and information based on the obfuscated data.

Anonymization schemes combine the two previous approaches by removing the identifying fields from the data, obfuscating the values in other fields, and adding randomness to the data all to guarantee a level of $k$-anonimity for each user in the dataset. Examples of this scheme in recent literature include \cite{6615356} and \cite{Casino2013490}.

Obfuscation-based techniques have the advantage of being scalable since the transformations usually only need to be applied to the data at the point of origin, after which the obfuscated data can be used directly. However, the security of these techniques is harder to prove since it relies on randomness and anonymity, and it is harder to prove that a clever inference attack might not be able to re-identify some of the users. Another concern for this family of techniques is the accuracy, since adding randomness to the data can lead to the loss of some key information that some CF algorithms might be able to benefit from in providing more accurate recommendations.

\subsubsection{Clustering-based techniques}\label{ssec:clustering}

Clustering-based techniques for PPCF rely on grouping the users into clusters or communities, then extracting a representation of that cluster and using it, providing anonymity for the users within each cluster. This can be used to protect against inference attacks by semi-honest or malicious adversaries, provided that the clusters are large enough and are chosen appropriately. It can also be used as a user profile privacy guarantee, if the user profile data is not stored in the system, and only the cluster information is used to provide recommendations.

Contributions under this category may overlap the other categories, often relying on cryptographic techniques  \cite{7044660,6570605,5066586,Ahmad2007} or hashing \cite{6406488} to extract the representation of each cluster and perform the CF calculations securely.

This approach has an obvious advantage in terms of scalability over the sole use of cryptographic techniques, since a reduced representation of the data is used instead of trying to encrypt the entire data set. However, care must be taken when implementing such a technique, to ensure that the accuracy of the results does not suffer due to this reduced representation.

\subsubsection{Other Approaches}\label{ssec:heuristic}

Some contributions in recent literature (e.g.: \cite{5616200,5474755,6785787,6680862,6682707,7095527}) resorted to the use of alternative approaches to provide a reasonable compromise that still maintains users' privacy without incurring the computation cost required by other methods. These approaches are often based on knowledge of the problem domain and rely on algorithms that are especially designed to leverage this knowledge.

In \cite{5616200}, the authors built a CF system based on expert opinions, in which the items are rated by domain experts instead of relying on user ratings, thus eliminating the privacy concerns altogether. Similarly, the authors in \cite{5474755} substituted item-similarity in place of user-similarity in their CF scheme, which removed the need for user profile data in the system and instead used the item data to calculate the recommendations. Another item-based approach was used in \cite{6785787} and \cite{7095527}. They used a distributed belief propagation approach that relies on statistical measures to provide the users with recommendations based on item-similarity without the need to store the user preferences in the system.

Alternative approaches to PPCF have the advantage of leveraging the problem domain knowledge to avoid the need for costly cryptographic operations or applying transformations to the data. However, they can potentially be of limited use since they rely on knowledge from specific domains. Item-based approaches may also sacrifice on recommendation accuracy since the users' preferences may contain key information that helps the CF system provide accurate and relevant recommendations.

\section{Conclusion}\label{sec:conclusion}

This survey examined the recent literature on PPCF recommendation systems from a broad perspective. The contributions were classified based on the vulnerability they address, then divided into a number of categories representing the type of approach used in each contribution. The different categories were discussed in terms of the different vulnerabilities they address, and a discussion of some of the potential advantages and disadvantages of each category was provided. Considering the rising popularity of CF systems and the equal rise of privacy awareness by users, this survey aims to assist researchers and practitioners interested in the development of practical and secure CF systems identify the most suitable approach for their particular problem based on the recent contributions in the field.

\section*{Acknowledgement}

Partial support for this research was received from the Missouri University of Science and Technology Intelligent Systems Center, the Mary K. Finley Missouri Endowment, the National Science Foundation, the Lifelong Learning Machines program from DARPA/Microsystems Technology Office, and the  Army  Research  Laboratory (ARL); and it was accomplished under Cooperative Agreement Number W911NF-18-2-0260. The views and conclusions contained  in  this  document  are  those  of  the  authors  and  should not  be  interpreted  as  representing  the  official  policies,  either expressed  or  implied,  of  the  Army  Research  Laboratory  or the  U.S.  Government.  The  U.S.  Government  is  authorized to reproduce and distribute reprints for Government purposes notwithstanding any copyright notation herein.

The authors would like to thank Emma Powell for her invaluable assistance with the organizational effort put into this survey. 

\IEEEtriggeratref{109}
\bibliographystyle{IEEEtran}
\bibliography{bibliography}

\end{document}